\documentclass[aps,prl,twocolumn,superscriptaddress,showpacs]{revtex4}
\usepackage{graphicx} %try this to include fig files
\usepackage{lineno}

\newcommand{\lapprox}{\stackrel{<}{_{\sim}}}

\begin{document}
\lefthyphenmin=2
\righthyphenmin=3

%new 060600
%\setlength{\baselineskip}{24pt}
%\preprint{FERMILAB-Pub-97/109-E}
%
% ======> Title of the paper goes here <====================
%
\title{
Observation of Exclusive Charmonium Production and $\gamma\gamma\rightarrow \mu^+\mu^-$ \\ 
in $p\bar{p}$ Collisions at $\sqrt{s}$ = 1.96 TeV
}

\affiliation{Institute of Physics, Academia Sinica, Taipei, Taiwan 11529, Republic of China} 
\affiliation{Argonne National Laboratory, Argonne, Illinois 60439} 
\affiliation{University of Athens, 157 71 Athens, Greece} 
\affiliation{Institut de Fisica d'Altes Energies, Universitat Autonoma de Barcelona, E-08193, Bellaterra (Barcelona), Spain} 
\affiliation{Baylor University, Waco, Texas  76798} 
\affiliation{Istituto Nazionale di Fisica Nucleare Bologna, $^v$University of Bologna, I-40127 Bologna, Italy} 
\affiliation{Brandeis University, Waltham, Massachusetts 02254} 
\affiliation{University of California, Davis, Davis, California  95616} 
\affiliation{University of California, Los Angeles, Los Angeles, California  90024} 
\affiliation{University of California, San Diego, La Jolla, California  92093} 
\affiliation{University of California, Santa Barbara, Santa Barbara, California 93106} 
\affiliation{Instituto de Fisica de Cantabria, CSIC-University of Cantabria, 39005 Santander, Spain} 
\affiliation{Carnegie Mellon University, Pittsburgh, PA  15213} 
\affiliation{Enrico Fermi Institute, University of Chicago, Chicago, Illinois 60637}
\affiliation{Comenius University, 842 48 Bratislava, Slovakia; Institute of Experimental Physics, 040 01 Kosice, Slovakia} 
\affiliation{Joint Institute for Nuclear Research, RU-141980 Dubna, Russia} 
\affiliation{Duke University, Durham, North Carolina  27708} 
\affiliation{Fermi National Accelerator Laboratory, Batavia, Illinois 60510} 
\affiliation{University of Florida, Gainesville, Florida  32611} 
\affiliation{Laboratori Nazionali di Frascati, Istituto Nazionale di Fisica Nucleare, I-00044 Frascati, Italy} 
\affiliation{University of Geneva, CH-1211 Geneva 4, Switzerland} 
\affiliation{Glasgow University, Glasgow G12 8QQ, United Kingdom} 
\affiliation{Harvard University, Cambridge, Massachusetts 02138} 
\affiliation{Division of High Energy Physics, Department of Physics, University of Helsinki and Helsinki Institute of Physics, FIN-00014, Helsinki, Finland} 
\affiliation{University of Illinois, Urbana, Illinois 61801} 
\affiliation{The Johns Hopkins University, Baltimore, Maryland 21218} 
\affiliation{Institut f\"{u}r Experimentelle Kernphysik, Universit\"{a}t Karlsruhe, 76128 Karlsruhe, Germany} 
\affiliation{Center for High Energy Physics: Kyungpook National University, Daegu 702-701, Korea; Seoul National University, Seoul 151-742, Korea; Sungkyunkwan University, Suwon 440-746, Korea; Korea Institute of Science and Technology Information, Daejeon, 305-806, Korea; Chonnam National University, Gwangju, 500-757, Korea} 
\affiliation{Ernest Orlando Lawrence Berkeley National Laboratory, Berkeley, California 94720} 
\affiliation{University of Liverpool, Liverpool L69 7ZE, United Kingdom} 
\affiliation{University College London, London WC1E 6BT, United Kingdom} 
\affiliation{Centro de Investigaciones Energeticas Medioambientales y Tecnologicas, E-28040 Madrid, Spain} 
\affiliation{Massachusetts Institute of Technology, Cambridge, Massachusetts  02139} 
\affiliation{Institute of Particle Physics: University of Alberta, Edmonton, Canada, T6G 2G7; McGill University, Montr\'{e}al, Qu\'{e}bec, Canada H3A~2T8; Simon Fraser University, Burnaby, British Columbia, Canada V5A~1S6; University of Toronto, Toronto, Ontario, Canada M5S~1A7; and TRIUMF, Vancouver, British Columbia, Canada V6T~2A3} 
\affiliation{University of Michigan, Ann Arbor, Michigan 48109} 
\affiliation{Michigan State University, East Lansing, Michigan  48824}
\affiliation{Institution for Theoretical and Experimental Physics, ITEP, Moscow 117259, Russia} 
\affiliation{University of New Mexico, Albuquerque, New Mexico 87131} 
\affiliation{Northwestern University, Evanston, Illinois  60208} 
\affiliation{The Ohio State University, Columbus, Ohio  43210} 
\affiliation{Okayama University, Okayama 700-8530, Japan} 
\affiliation{Osaka City University, Osaka 588, Japan} 
\affiliation{University of Oxford, Oxford OX1 3RH, United Kingdom} 
\affiliation{Istituto Nazionale di Fisica Nucleare, Sezione di Padova-Trento, $^w$University of Padova, I-35131 Padova, Italy} 
\affiliation{LPNHE, Universite Pierre et Marie Curie/IN2P3-CNRS, UMR7585, Paris, F-75252 France} 
\affiliation{University of Pennsylvania, Philadelphia, Pennsylvania 19104}
\affiliation{Istituto Nazionale di Fisica Nucleare Pisa, $^x$University of Pisa, $^y$University of Siena and $^z$Scuola Normale Superiore, I-56127 Pisa, Italy} 
\affiliation{University of Pittsburgh, Pittsburgh, Pennsylvania 15260} 
\affiliation{Purdue University, West Lafayette, Indiana 47907} 
\affiliation{University of Rochester, Rochester, New York 14627} 
\affiliation{The Rockefeller University, New York, New York 10021} 
\affiliation{Istituto Nazionale di Fisica Nucleare, Sezione di Roma 1, $^{aa}$Sapienza Universit\`{a} di Roma, I-00185 Roma, Italy} 
\affiliation{Rutgers University, Piscataway, New Jersey 08855} 
\affiliation{Texas A\&M University, College Station, Texas 77843} 
\affiliation{Istituto Nazionale di Fisica Nucleare Trieste/Udine, I-34100 Trieste, $^{bb}$University of Trieste/Udine, I-33100 Udine, Italy} 
\affiliation{University of Tsukuba, Tsukuba, Ibaraki 305, Japan} 
\affiliation{Tufts University, Medford, Massachusetts 02155} 
\affiliation{Waseda University, Tokyo 169, Japan} 
\affiliation{Wayne State University, Detroit, Michigan  48201} 
\affiliation{University of Wisconsin, Madison, Wisconsin 53706} 
\affiliation{Yale University, New Haven, Connecticut 06520} 
\author{T.~Aaltonen}
\affiliation{Division of High Energy Physics, Department of Physics, University of Helsinki and Helsinki Institute of Physics, FIN-00014, Helsinki, Finland}
\author{J.~Adelman}
\affiliation{Enrico Fermi Institute, University of Chicago, Chicago, Illinois 60637}
\author{T.~Akimoto}
\affiliation{University of Tsukuba, Tsukuba, Ibaraki 305, Japan}
\author{M.G.~Albrow}
\affiliation{Fermi National Accelerator Laboratory, Batavia, Illinois 60510}
\author{B.~\'{A}lvarez~Gonz\'{a}lez$^q$}
\affiliation{Instituto de Fisica de Cantabria, CSIC-University of Cantabria, 39005 Santander, Spain}
\author{S.~Amerio$^w$}
\affiliation{Istituto Nazionale di Fisica Nucleare, Sezione di Padova-Trento, $^w$University of Padova, I-35131 Padova, Italy} 
\author{D.~Amidei}
\affiliation{University of Michigan, Ann Arbor, Michigan 48109}
\author{A.~Anastassov}
\affiliation{Northwestern University, Evanston, Illinois  60208}
\author{A.~Annovi}
\affiliation{Laboratori Nazionali di Frascati, Istituto Nazionale di Fisica Nucleare, I-00044 Frascati, Italy}
\author{J.~Antos}
\affiliation{Comenius University, 842 48 Bratislava, Slovakia; Institute of Experimental Physics, 040 01 Kosice, Slovakia}
% ...

\collaboration{CDF Collaboration}
\noaffiliation

\begin{abstract}
 In CDF we have observed the reactions $p+\bar{p} \rightarrow p + X +
 \bar{p}$, with $X$ being a centrally produced $J/\psi,\psi(2S)$ or $\chi_{c0}$,
and $\gamma\gamma\rightarrow \mu^+\mu^-$, in $p\bar{p}$ 
collisions at $\sqrt{s}$ = 1.96 TeV. 
The event signature requires two oppositely charged central muons and 
either no other particles, or one additional photon, detected.
Exclusive vector meson production is as expected for elastic photoproduction,
$\gamma + p \rightarrow J/\psi(\psi(2S)) + p$, observed here for the first 
time in hadron-hadron collisions. 
We also observe exclusive $\chi_{c0} \rightarrow J/\psi+\gamma$. 
The cross sections $\frac{d\sigma}{dy}|_{y=0}$ for  
$J/\psi,\psi(2S)$ and $\chi_{c0}$
are 3.92$\pm$0.25(stat)$\pm$0.52(syst) nb, 0.53$\pm$0.09(stat)$\pm$0.10(syst) nb, 
and 76$\pm$10(stat)$\pm$10(syst) nb respectively, and the continuum
is consistent with QED. We put an upper 
limit on the cross section for odderon exchange in exclusive
$J/\psi$ production.

\end{abstract}
\maketitle

In central exclusive production processes, $p+\bar{p}\rightarrow p+X+\bar{p}$, the colliding 
hadrons emerge intact with small transverse momenta, $p_T$~\cite{kine}, and the produced state
$X$ is in the
central region, with small rapidity $|y|$, and is fully measured. 
If regions of rapidity exceeding about 5 units are devoid of particles,
only photon and pomeron~\cite{pbooks}, $I\!\!P$,
exchanges are significant, where $I\!\!P$ consists mostly of two gluons in a color singlet state with charge parity
$C = +1$.
Odderon, $O$, exchange, with 3 gluons in a $C = -1$ state\cite{odd,kmr,ewerz}, is allowed
in $p\bar{p}$, but not $ep$, collisions, and would appear as an enhancement in exclusive $J/\psi$ and $\psi(2S)$ production in
$p\bar{p}$ compared to the expectation from pure photoproduction in $ep$.
Using the CDF II detector at the Fermilab Tevatron,
we previously observed~\cite{exclee} $p+\bar{p} \rightarrow p+ e^+e^- + \bar{p}$
in agreement with QED, and found candidates~\cite{exclgg} for $p+\bar{p} \rightarrow p+ \gamma\gamma + \bar{p}$ consistent
with QCD expectations~\cite{KMRgg}. In this paper we
report measurements of exclusive dimuon production, $X=\mu^+\mu^-$, with M$_{\mu\mu} \in [3.0,4.0]$ GeV/c$^2$, 
directly (QED, Fig.1a) or from photoproduced $J/\psi(3097)$ or $\psi(2S)(3686)$ (Fig.1b)
decay, and $\chi_{c0}(3415) \rightarrow J/\psi + \gamma \rightarrow \mu^+\mu^-\gamma$ (Fig.1c). 
Lower masses were excluded by muon range, and higher masses by trigger rate limitations.
Exclusive photoproduction of
vector mesons has been measured in $ep$
collisons at HERA~\cite{hera}, but not previously observed in hadron-hadron collisions.
The theoretical uncertainty on the QED cross section is $<$ 0.3\%; 
this process is distinct from Drell-Yan ($q\bar{q}\rightarrow \mu^+\mu^-$), which is negligible in this 
regime.

At the Large Hadron Collider, LHC, 
in $pp$ collisions with $\sqrt{s}$ = 10-14 TeV, central exclusive production of states such as
$X = H$ and $W^+W^-$, where H is a Higgs boson, are allowed~\cite{fp420}. 
Apart from their intrinsic interest, our measurements
confirm the viability of the proposed LHC studies. The $p+\chi_{c0} + \bar{p}$ (Fig.1c) 
and $p+H+p$ (as Fig.1c but with a top quark loop) cross
sections are related~\cite{kmr04}, and $p+ \mu^+\mu^- + p$ can be used to calibrate forward proton spectrometers.

\begin{figure}[t]
\vspace{1.5cm}
  \includegraphics{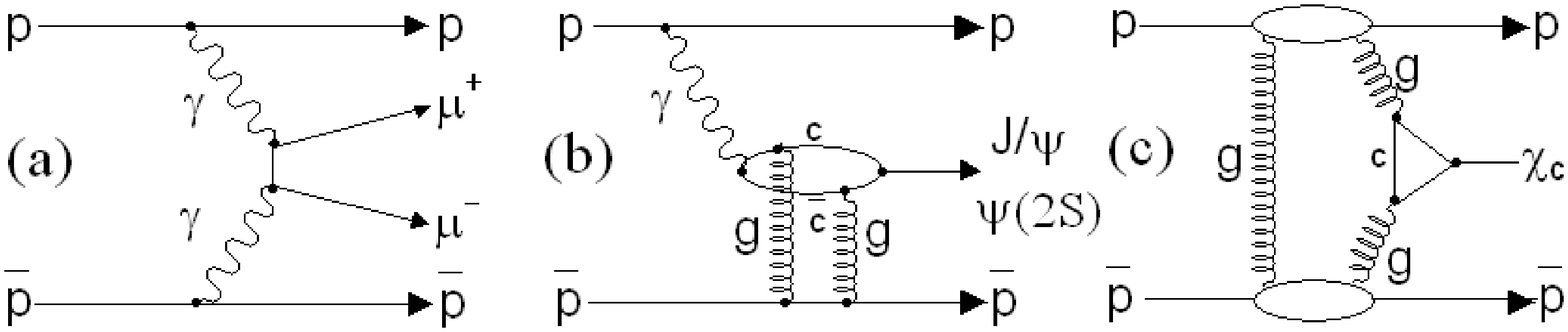}
    \vspace{0.5cm}
    \caption{Feynman diagrams for (a) $\gamma\gamma\rightarrow \mu^+\mu^-$, (b) $\gamma I\!\!P \rightarrow 
    J/\psi (\psi(2S))$, and
    (c) $I\!\!P I\!\!P\rightarrow\chi_c$, with the 2-gluon exchange forming a pomeron.
\label{fig:emet} }
\end{figure}

We used $\bar{p} p$ collision data at $\sqrt{s}$ = 1.96 TeV with an integrated luminosity $L$ = 1.48 fb$^{-1}$ 
delivered to the CDF II detector.
This is a general purpose detector described elsewhere~\cite{cdf}.
Surrounding the collision region is a tracking system consisting of
silicon microstrip detectors and a cylindrical drift
chamber (COT) in a 1.4~Tesla solenoidal field.  The tracking system has $\approx$ 100\% efficiency
for reconstructing isolated tracks with $p_T \geq$ 1 GeV/c and $|\eta|<0.6$~\cite{kine}.  
A barrel of 216 time-of-flight counters (ToF) outside the COT is surrounded by
calorimeters with separate electromagnetic (EM) and hadronic sections covering the
range $|\eta|<3.6$. Drift
chambers outside the calorimeters were used to measure muons with $|\eta| <$ 0.6~\cite{muondc}.
The regions $3.6<|\eta|<5.2$ are covered by
lead-liquid scintillator calorimeters~\cite{miniplug}. Gas
Cherenkov counters covering 3.7$<|\eta|<$4.7 determined the luminosity with a 6\%
uncertainty~\cite{clc}.
We did not have detectors able to measure the forward $p$ and $\bar{p}$, but beam
shower scintillation counters (BSC1-BSC3), located
along the beampipe, detected products of $p(\bar{p})$ fragmentation, such as $p\rightarrow p\pi\pi$, with $|\eta| <$ 7.4.

The level 1 trigger required at least one muon track with $p_T >$ 1.4 GeV/c and no signal in 
BSC1 (5.4 $\lapprox |\eta| \lapprox$ 5.9), and a higher level trigger required a second track with opposite charge.
The offline event selection followed closely that described in Ref.~\cite{exclee}, where we observed exclusive $e^+e^-$
production. We required two oppositely charged muon tracks, each with $p_T >$ 1.4 GeV/c and $|\eta|<$ 0.6, 
accompanied by either
(a) no other particles in the event, or (b) only one additional 
EM shower with $E_T^{EM} > 80$ MeV and $|\eta| <$ 2.1. Condition (a) defines an exclusive dimuon event. 
The exclusivity efficiency \mbox{$\varepsilon_{\rm exc}$} is the probability that the exclusive requirement is not
spoiled by another inelastic interaction in the same bunch crossing, or by noise in a detector element. 
This efficiency was measured~\cite{exclee}
as the fraction of bunch crossing triggers that pass the exclusivity requirement (a).
We found \mbox{$\varepsilon_{\rm exc}= 0.093$} with negligible uncertainty.
% a factor $\approx$ 2 lower than the probability of having no additional interactions. 
The product \mbox{$\varepsilon_{\rm exc}$}$\times L$ = $L_{eff}$ =139 $\pm$ 8 pb$^{-1}$ was the
effective luminosity for single interactions.

  \begin{figure}[t]
 \vspace{5.5cm}
  \includegraphics{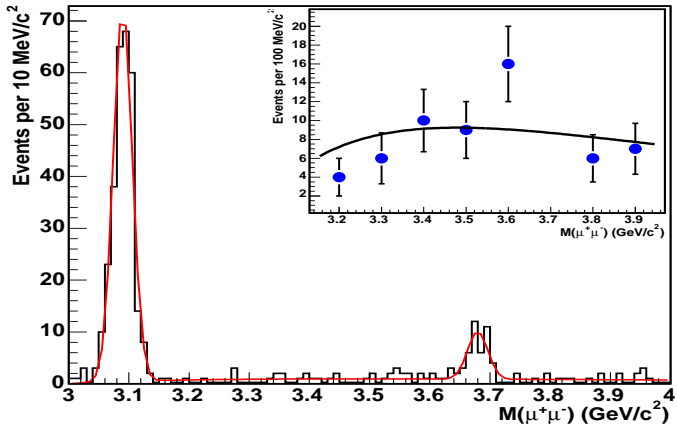}
   \vspace{0.5cm}
   \caption{Mass $M_{\mu\mu}$ distribution of 402 exclusive events, with no EM shower, (histogram) 
   together with a fit to two Gaussians for the $J/\psi$
    and $\psi(2S)$, and a QED continuum. All three shapes are predetermined, with only the normalizations floating. Inset: Data above the 
    $J/\psi$ and excluding $3.65 < M_{\mu\mu} < 3.75$ GeV/c$^2$ ($\psi(2S)$) with the fit to the QED spectrum times
    acceptance (statistical uncertainties only).
\label{fig:mass} }
\end{figure}

  After these selections, cosmic rays were the main background. 
  They were all rejected, with no significant loss of real events, by timing requirements in the ToF counters
  and by requiring the 3D opening angle between the muon tracks to be $\Delta\theta_{3D}(\mu\mu) < 3.0$ rad.
   Within a fiducial kinematic region (FKR): [$|\eta(\mu)|<$ 0.6, and M$_{\mu\mu} \in [3.0,4.0]$ GeV/c$^2$], there are 402 events 
   with no EM shower. The $M_{\mu\mu}$ spectrum is 
   shown in Fig.2. The
   $J/\psi$ and $\psi(2S)$ are prominent, together with a continuum. 
   The spectrum is well fitted by two Gaussians with expected masses and widths (dominated by
   the resolution) and a continuum whose shape is given by the product of the QED spectrum ($\gamma\gamma\rightarrow
   \mu^+\mu^-$), acceptance, and
   efficiency, as shown in Fig.2(inset). The numbers of events from the fit are given in 
   Table I, with statistical uncertainties. 
   The numbers given in Table I for backgrounds, acceptances,
   and efficiencies show systematic uncertainties estimated by varying parameters within acceptable bounds.

\begin{table}[htbp]
\caption{Numbers of events fitted to classes $J/\psi,\psi(2S)$, $\chi_{c0}$ and QED. Backgrounds are 
given as percentages of the fit events, and
efficiencies are to be applied to the events without background. The stated branching fraction $\mathcal{B}$ 
for the $\chi_{c0}$ is the product of the
$\chi_{c0} \rightarrow J/\psi + \gamma$ and $J/\psi \rightarrow \mu^+\mu^-$ branching fractions~\cite{pdg}.
For events(fit) the uncertainty is only statistical; all other uncertainties are purely 
systematic except when both are given.
The cross sections include a 6\% luminosity uncertainty.}
\begin{center}
\begin{tabular}{ccccc}	
\hline
\hline

Class & $J/\psi$ & $\psi(2S)$ & $\chi_{c0}(1P)$ & $\gamma\gamma \rightarrow \mu^+\mu^-$  \\ \hline
Acceptances: &              &           &           &         \\
Detector(\%)   & 18.8$\pm$2.0  & 54$\pm$3 & 19$\pm$2 & 41.8$\pm$1.5  \\
Efficiencies:              &            &           &      \\
$\mu$-quality(\%) & 33.4$\pm$1.7  & 45$\pm$6  & 33$\pm$2 & 41.8$\pm$2.3 \\
Photon(\%)& -  & -  & 83$\pm$4 & - \\
Events(fit)      & 286$\pm$17   & 39$\pm$7   & 65$\pm$8 & 77$\pm$9\\
Backgrounds: &         &  &      &     \\
Fragmention(\%) & 9$\pm$2   & 9$\pm$2         & 11$\pm$2 & 8$\pm$2 \\
Non-excl.(\%) & 3$\pm$3     & 3$\pm$3     & 3$\pm$3 & 9$\pm$5 \\
$\chi_{c0} (\%)$    & 4.0$\pm$1.6        &    -          &    -         &  - \\
%Events-B/G       & 243$\pm$21     & 34$\pm$7    & 65$\pm$10    & 56$\pm$8 \\
$\mathcal{B} \rightarrow\mu^+\mu^-$(\%)& 5.93$\pm$0.06 & 0.75$\pm$0.08 & 0.076 &  -  \\
      &    &    &      $\pm$0.007 &  \\
$\mathcal{B}.\sigma_{FKR}$(pb)& 28.4$\pm$4.5 & 1.02$\pm$0.26      & 8.0$\pm$1.3 & 2.7\\
$\frac{d\sigma}{dy}|_{y=0}$(nb) & 3.92 & 0.53          & 76  &    -   \\
      & $\pm$0.25(stat)     & $\pm$0.09(stat)   &      $\pm$10(stat) &  $\pm$0.3(stat)  \\
      &  $\pm$0.52(syst)   & $\pm$0.10(syst)    &     $\pm$10(syst)  &   $\pm$0.4(syst)       \\
\hline
\hline
\end{tabular}
\end{center}
\label{tab:events} 
\end{table}
   
    Backgrounds to exclusive $\mu^+\mu^-$ events are (see Table I) (a) proton fragmentation, if the products are not detected
 in the forward detectors, (b) for the $J/\psi$, $\chi_{c0}$ events with a photon that did not give an EM 
 shower above 80 MeV, and (c) events with 
 some other particle not
 detected. The probability of a $p$ or $\bar{p}$ fragmenting at the $p\gamma p(p^*)$ 
 vertex was calculated with the \textsc{lpair} 
 Monte Carlo (MC) simulation~\cite{lpair} to be 0.17$\pm$0.02(syst), and the probability
 that all the fragmentation products have $|\eta| >$ 7.4 to be 0.14$\pm$0.02(syst). If a proton fragments,
 the decay products may not be detected through BSC inefficiency, estimated from data to be 0.08$\pm$0.01. 
 The fragmention probability at the $p I\!\!P p(p^*)$ vertex was taken from the ratio of single diffractive 
 fragmentation to elastic scattering at the Tevatron~\cite{cdfdiff} to be 0.24$\pm$0.05. 
  
We compared the kinematics of the muons, e.g. $p_T(\mu^+\mu^-)$ and
 $\Delta\phi_{\mu\mu}$, with simulations for the three 
 classes: $J/\psi, \psi(2S)$~\cite{nystrand}, and QED~\cite{lpair}
 with M$_{\mu\mu} \in  [3.2,3.6]\&[3.8,4.0]$ GeV/c$^2$ to exclude the $J/\psi$ and $\psi(2S)$. The distributions
  agree well with the simulations; the few events that are outside expectations are taken to be non-exclusive background. 
 Figure 3 shows the distributions of $p_T(\mu^+\mu^-)$. As expected, $\langle p_T \rangle$ is smaller for the QED process, and the data agree 
 well with \textsc{starlight}~\cite{nystrand}, apart from two events with $p_T >$ 0.8 GeV/c where no events are expected. 
 Comparing data with \textsc{lpair} we estimate that the non-exclusive background is (9$\pm$5)\% of the 
observed(QED) events.
 The $\psi(2S)$ data are well
 fitted by the \textsc{starlight} photoproduction simulation\cite{nystrand}. The distribution of $p_T(J/\psi)$ is well 
 fitted by \textsc{starlight}, apart from five events
 with $p_T(J/\psi) >$ 1.4 GeV/c (Fig. 3b). These could be due to non-exclusive background, some $\chi_{c0}$ radiative decays with
 an undetected photon, or an odderon component. 

   \begin{figure}[t]
\vspace{1.0cm}
  \includegraphics{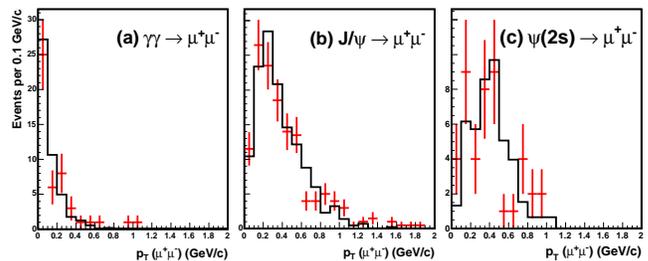}
  \vspace{2.0cm}
   \caption{$p_T$ distribution of $\mu^+\mu^-$ (points with statistical error bars) for (a) QED, $M_{\mu\mu} \in$ [3.2,3.6]+[3.8,4.0]
    GeV/c$^2$, (b) $J/\psi$, and (c) $\psi(2S)$. The MC predictions (with no background) are shown by the 
    histograms, normalized to the data. 
\label{fig:ptdist} }
\end{figure}

 To measure $\chi_{c0}$ production we required one EM shower with $E_T^{EM} > 80$ MeV in addition to the two muons; if two 
 adjacent towers had enough energy they were combined. There are 65 events in the $J/\psi$ peak 
 and eight continuum events; these are likely to be $\gamma\gamma \rightarrow \mu^+\mu^-$ with a
 bremsstrahlung. We interpret the 65 events as $\chi_{c0} \rightarrow J/\psi+\gamma$ production and decay. 
  The distribution of the
 mass formed from the $J/\psi$ and the EM shower energy, while broad, has a mean value equal to the $\chi_{c0}$ mass. 
  The  $E_T^{EM}$ spectrum is well fitted by an empirical function which extrapolates to 
  only 3.6$\pm$1.3(syst) $\chi_{c0}$ candidates with showers below 80 MeV.  The $p_T(J/\psi)$ and $\Delta\phi_{\mu\mu}$ distributions for the events with an $E_T^{EM}$ signal are consistent
  with all these $J/\psi$ being from $\chi_{c0}$ decay, as simulated by \textsc{chicmc}~\cite{chicmc}.  
  Additional photon inefficiency comes from conversion in material, 7$\pm$2\%, and dead regions of the calorimeter,
  5.0$\pm$2.5\%, giving a total inefficiency 17$\pm$4\%, which gives a background to exclusive $J/\psi$ of 
  4.0$\pm$1.6\% (all errors syst.).

We calculated acceptances and efficiencies using the \textsc{lpair}~\cite{lpair} 
and \textsc{starlight}~\cite{nystrand} MC generators
 for QED, $J/\psi$ and $\psi(2S)$, and \textsc{chicmc}~\cite{chicmc} for $\chi_{c0}$ production.
Generated events were passed through a \textsc{geant}-based~\cite{geant} simulation of the CDF detector. 
The trigger efficiency for muons rose steeply between
1.4 GeV/c and 1.5 GeV/c, where it exceeded 90\%. As we triggered on one muon, the trigger efficiency 
for events with two muons was $>99$\% for $M_{\mu\mu}>$ 3 GeV/c$^2$. 
 
Figure 2(inset) shows the subset of the Fig.2 data above 3.15 GeV/c$^2$ (to exclude the $J/\psi$), excluding the bin 
3.65-3.75 GeV/c$^2$ which contains the $\psi(2S)$. The curve shows the product of the QED spectrum and acceptance $\times$ efficiency, $A\varepsilon$, 
with only the normalization floating, from the 3-component fit to the full spectrum. 
The continuum data agrees with the QED expectation.
The integral from 3 GeV/c$^2$ to 4 GeV/c$^2$ is 77$\pm$9(stat) events, and after correcting for backgrounds and efficiencies (Table I), 
the measured cross section for QED events with 
$|\eta(\mu^\pm)| < 0.6$ and M$_{\mu\mu} \in [3.0,4.0]$ GeV/c$^2$ is $\sigma$ =
2.7$\pm$0.3(stat)$\pm$0.4(syst) pb, 
in agreement with the QED prediction 2.18$\pm$0.01 pb~\cite{lpair}.

For the prompt $J/\psi$ and $\psi(2S)$ cross sections we took the number of events from the Gaussian fits, subtracted 
backgrounds and corrected for $A\varepsilon$ to obtain $\mathcal{B} \times \sigma_{FKR}$ for 
both muons in the fiducial kinematic region (see Table I). To obtain $\frac{d\sigma}{dy}|_{y=0}$ from $\sigma_{FKR}$ we used the 
\textsc{starlight} MC, which gives the ratio of these two cross sections for each resonance, and divided by the branching
fractions $\mathcal{B}$.
We found $\frac{d\sigma}{dy}|_{y=0}(J/\psi) = 3.92\pm0.25$(stat)$\pm$0.52(syst) nb. This agrees with the 
predictions 2.7$^{+0.6}_{-0.2}$ nb~\cite{nystrand} and 3.4$\pm$0.4
nb~\cite{motyka} among others~\cite{scha,goncalves}. We found 
$\frac{d\sigma}{dy}|_{y=0}(\psi(2S)) = 0.53\pm0.09$(stat)$\pm$0.10(syst) nb compared with a prediction~\cite{nystrand} 0.46$^{+0.11}_{-0.04}$ nb. The ratio $R = \frac{\psi(2S)}{J/\psi}$ = 0.14$\pm$0.05 is in 
agreement with the HERA value~\cite{hera} $R = 0.166\pm$0.012 at similar $\sqrt{s(\gamma p)}$. 

After correcting the 65 $\chi_{c0}$ candidates for backgrounds and efficiencies,
and applying the branching 
fraction $\mathcal{B}(\chi_{c0} \rightarrow J/\psi+\gamma)$ =
0.0128$\pm$0.0011~\cite{pdg} we found $\frac{d\sigma}{dy}|_{y=0}(\chi_{c0}) = 76\pm
10$(stat)$\pm$10(syst) nb.
The $\chi_{c2}(3556)$
may be present, although it is strongly suppressed by the $J_z=0$ 
rule~\cite{kmr04} and is forbidden at 0$^\circ$ scattering angle.
Exclusive $gg\rightarrow \chi_{c1}(3511),J^{PC} = 1^{++}$ is forbidden by the Landau-Yang theorem, but may occur
with off-shell gluons~\cite{pasechnik}. It is nevertheless forbidden by symmetry arguments~\cite{kkmr} 
when both $p$ and $\bar{p}$ scatter at 0$^\circ$.
Because of the limited $M(J/\psi+\gamma)$ resolution we cannot distinguish these states; we assume 
$\chi_{c1}$ and $\chi_{c2}$ to be
negligible. If several states $\chi_{ci}$ are present, $\sum \mathcal{B}_i \sigma_{i,FKR}$ =
8.0$\pm$0.9(stat)$\pm$0.9(syst) pb.
Theoretical predictions have 
large (often unstated) uncertainties, but are compatible with our measurement. Ref.~\cite{kmr04} 
predicted $\frac{d\sigma}{dy}|_{y=0}(\chi_{c0})$ = 130 nb;
however the P.D.G value~\cite{pdg} of the $\chi_c$ width has since been reduced by a factor 1.45, correcting their
prediction to 90 nb. Yuan~\cite{yuan} predicted 160 nb (again the factor $\frac{1}{1.45}$ should be applied) and
Bzdak~\cite{bzdak} 45 nb. 

If the $J/\psi$ and $\psi(2S)$ cross sections were larger than expected for photoproduction, it would be evidence for odderon
exchange.
Taking a theoretical value of $\frac{d\sigma}{dy}|_{y=0}(J/\psi)= 3.0\pm$0.3 nb 
for photoproduction, 
compatible with the predictions,
we give a 95\% C.L. upper limit $\frac{d\sigma}{dy}|_{y=0}(J/\psi) < 2.3$ nb for odderon 
exchange ($O I\!\!P \rightarrow J/\psi$). Bzdak et al.~\cite{bmsc} predicted the ratio of odderon:photon exchange in
$J/\psi$ production to be 0.3 - 0.6, consistent with our limit.

In conclusion we have observed, for the first time in hadron-hadron collisions, exclusive 
photoproduction of $J/\psi$ and $\psi(2S)$, exclusive 
double pomeron production of $\chi_{c0}$, and the QED process $\gamma\gamma
\rightarrow \mu^+\mu^-$. 
 The photoproduction process has previously been studied in
$ep$ collisions at HERA, with similar kinematics ($\sqrt{s(\gamma p)} \approx$ 100 GeV) and the cross sections 
are in agreement.
We put an upper limit on an odderon contribution to exclusive $J/\psi$
production. Our observation of exclusive $\chi_{c0}$ production 
implies that exclusive Higgs boson production should occur at the LHC~\cite{fp420} and imposes 
constraints on the $p+p\rightarrow p+H+p$ cross section.

%For PRD change to:
%\begin{acknowledgments}
%check for the latest@
%http://www-cdf.fnal.gov/internal/people/links/CarolPicciolo/standardack.html
\section{Acknowledgments}
We thank the Fermilab staff and the technical staffs
of the participating institutions for their vital contributions.
This work was supported by the U.S. Department of Energy and National
Science Foundation; the Italian Istituto Nazionale di Fisica Nucleare;
the Ministry of Education, Culture, Sports, Science and Technology
of Japan; the Natural Sciences and Engineering Research Council
of Canada; the National Science Council of the Republic of China;
the Swiss National Science Foundation; the A.P. Sloan Foundation;
the Bundesministerium f\"{u}r Bildung und Forschung, Germany;
the Korean Science and Engineering Foundation and the Korean
Research Foundation; 
the Science and Technology Facilities Council and the Royal Society, UK;
the Institut National de Physique Nucleaire et Physique des Particules/CNRS;
the Russian Foundation for
Basic Research; the Ministerio de Ciencia e Innovaci\'{o}n, and Programa 
Consolider-Ingenio 2010, Spain; the Slovak R\&D Agency; and the Academy of Finland. 
%\end{acknowledgments}


\begin{thebibliography}{99} 
\bibitem{kine} A cylindrical coordinate system is used with the $z$-axis along the 
proton beam direction; $\theta$ is the polar angle and $\phi$ is the azimuthal angle.
Transverse momentum is 
$p_T =|p| \sin\theta$, and transverse energy is $E_T = E \sin\theta$ where $E$ is the energy. Pseudorapidity is
$\eta = -\ln(\tan\frac{\theta}{2})$, and for the charmonium 
states we use longitudinal rapidity $y = -\ln \frac{E+p_z}{E-p_z}$.
\bibitem{pbooks} See e.g. J.R.Forshaw and D.A.Ross, Quantum Chromodynamics and the Pomeron, Cambridge University Press
(1997); S.Donnachie \emph{et al.}, Pomeron Physics and QCD, Cambridge University Press (2002).
\bibitem{odd} A.Sch\"{a}fer, L.Mankiewicz and O.Nachtmann,
Phys.Lett. B {\bf 272}, 419 (1991).
\bibitem{kmr} V.A.Khoze, A.D.Martin and M.G.Ryskin, Eur. Phys. J. C {\bf 24}, 459 (2002).
\bibitem{ewerz} C.Ewerz, The Odderon in Quantum Chromodynamics, hep-ph/0306137.
\bibitem{exclee} A.Abulencia \textit{et al.} (CDF Collaboration), Phys. Rev. Lett. {\bf 98}, 112001 (2007).
\bibitem{exclgg} A.Abulencia \textit{et al.} (CDF Collaboration), Phys. Rev. Lett. {\bf 99}, 242002 (2007).
\bibitem{KMRgg} V.A.Khoze, A.D.Martin, M.G.Ryskin and W.J.Stirling, Eur. Phys. J. C {\bf 38}, 475 (2005).
\bibitem{hera} See e.g. H.Jung, Acta Phys. Polon. Supp. {\bf 1}, 531 (2008), and refs. therein.
\bibitem{fp420} M.G.Albrow \textit{et al.}, arXiv:0806.0302 [hep-ex], submitted to J. Inst.
\bibitem{kmr04} V.A.Khoze, A.D.Martin, M.G.Ryskin and W.J.Stirling, Eur. Phys. J. C {\bf 35}, 211 (2004);
V.A.Khoze, A.D.Martin and M.G.Ryskin, Eur. Phys. J. C {\bf 19}, 477 (2001) [Erratum-ibid
{\bf C20}, 599 (2001)], and private communication.
\bibitem{cdf} D.Acosta \textit{et al.} (CDF Collaboration), Phys. Rev. D {\bf 71}, 032001 (2005) 
and refs. therein.
\bibitem{muondc} G.Ascoli \textit{et al.}, Nucl. Instrum. Methods A {\bf 268}, 33 (1988).
\bibitem{miniplug} M.Gallinaro \textit{et al.}, IEEE Trans. Nucl. Sci. {\bf 52}, 879 (2005).
\bibitem{clc} D.Acosta \textit{et al.}, Nucl. Instrum. Methods A {\bf 494}, 57 (2002).
\bibitem{pdg} C.Amsler \textit{et al.}, Review of Particle Physics, Phys.Lett. B {\bf 667}, 1 (2008).
\bibitem{lpair} J.A.M.Vermaseren, Nucl. Phys. B {\bf 229}, 347 (1983); S.P.Baranov \textit{et al.}, \textsc{lpair}, 
Proc. Physics at HERA, 1478 (1991).
\bibitem{cdfdiff} F.Abe \textit{et al.} (CDF Collaboration), Phys. Rev. D {\bf 50}, 5518 (1994); ibid. p.5535.
\bibitem{nystrand} S.Klein and J.Nystrand, \textsc{starlight}, Phys. Rev. Lett. \textbf{92}:142003 (2004), and private communication.
\bibitem{chicmc} W.J.Stirling, \textsc{chicmc}, private communication.
\bibitem{geant} \textsc{geant}, CERN Program Library Long Writeup W5013, Nucl. Instrum. Methods A {\bf 506}, 
250 (2003).
\bibitem{motyka} L.Motyka and G.Watt, Phys. Rev. D {\bf 78}:014023 (2008).
\bibitem{scha} W.Sch\"{a}fer and A.Szczurek, Phys. Rev. D {\bf 76}, 094014 (2007).
\bibitem{goncalves} V.P.Goncalves and M.V.T.Machado, Eur. Phys. J. C {\bf 40}, 519 (2005).
\bibitem{pasechnik} R.S.Pasechnik, A.Szczurek, and O.V.Teryaev, arXiv:0901.4187 [hep-ph].
\bibitem{kkmr} A.D.Kaidalov, V.A.Khoze, A.D.Martin and M.G.Ryskin, Eur. Phys. J. C {\bf 31}, 387 (2003).
\bibitem{yuan} F.Yuan, Phys.Lett. B {\bf 510}, 155 (2001).
\bibitem{bzdak} A.Bzdak, Phys.Lett. B {\bf 619}, 288 (2005).
\bibitem{bmsc} A.Bzdak, L.Motyka, L.Szymanowski and J-R.Cudell, Phys. Rev. D {\bf 75}:094023 (2007).

\end{thebibliography}
\end{document}